# The spectra of accretion discs in low-mass X-ray binaries

R. R. Ross[1,2] and A. C. Fabian[2]

[1] *Physics Department, College of the Holy Cross, Worcester, MA 01610, USA (Internet: ross@hcacad.holycross.edu)*
[2] *Institute of Astronomy, Madingley Road, Cambridge CB3 0HA (Internet: acf@ast.cam.ac.uk)*

1 November 1995

**ABSTRACT**
We present self-consistent models for the radiative transfer in Shakura-Sunyaev accretion discs in bright low-mass X-ray binaries (LMXB). Our calculations include the full effects of incoherent Compton scattering and the vertical temperature structure within the disc, as well as the effects of Doppler blurring and gravitational redshift. We find that the observed X-ray spectra are well fit by exponentially cutoff power-law models. The difference between the observed total spectrum and our calculated disc spectrum should reveal the spectrum of the disc/neutron star boundary layer and other emitting regions considered to be present in LMXB.

## 1 INTRODUCTION

The energy spectra of low-mass X-ray binaries (LMXB) are observed to be approximately exponential in shape with characteristic temperatures of a few keV. As the magnetic field of the central neutron star is weak, there is much emission from the accretion disc, as well as from a boundary layer between the disc and neutron star. Traditionally the disc emission has been modelled by summing blackbody or modified blackbody spectra assuming that there is a uniform temperature at each disc radius (Mitsuda et al. 1984; White et al. 1986; Czerny, Czerny & Grindlay 1986; Taam & Meszaros 1987; White, Stella & Parmar 1988; Nannurelli & Stella 1989). However, such spectra must be poor approximations to a real disc in which there is a vertical temperature gradient at each radius (and photons of different energies emerge from different depths) and in which the effects of *incoherent* Compton scattering have been shown to be very important.

Recenty, Shimura & Takahara (1995) have presented self-consistent calculations for the radiative transfer and vertical structure in accretion discs of black hole candidates. They have extended the method that we used (Ross, Fabian & Mineshige 1992) for treating the radiative transfer in AGN accretion discs to include the vertical density structure as well as temperature structure within the disc. One of the models that Shimura & Takahara have presented was for a 1.4-$M_\odot$ central object, appropriate for a LMXB accretion disc. However, their total disc spectra were simple sums of locally emergent spectra; they have not been corrected for gravitational redshift and Doppler blurring to correspond to the spectra that would actually be observed.

Here we present self-consistent models for the radiative transfer in Shakura-Sunyaev (1973) accretion discs about a 1.4-$M_\odot$ neutron star, including first-order corrections for gravitational redshift and Doppler blurring. We make the standard assumption of a uniform density structure at each radius within the disc. Shimura & Takahara (1993) have found that, at high accretion rates, the density in the inner accretion disc is nearly uniform with depth (except in the outermost Thomson mean free path). We give simple fits to our final models in terms of exponentially cutoff power-law spectra. We hope that the fits will be useful for the interpretation of existing (e.g., *Ginga*) and future (e.g., *XTE*) data from LMXB.

This work is part of an ongoing project to calculate self-consistent spectra of accretion discs. We plan to extend this work in the future to include the effects of heavy-element atomic processes and external illumination of the surface for LMXB accretion discs.

## 2 METHOD

We treat the radiative transfer in the inner, radiation-dominated portions of a LMXB accretion disc. A geometrically thin, $\alpha$-viscosity disc is assumed as described by Shakura & Sunyaev (1973). The equations for the structure of the disc have been summarized by Ross, Fabian & Mineshige (1992). The central neutron star has mass $M = 1.4 M_\odot$. The accreting plasma is taken to consist of fully-ionized hydrogen and helium, with the relative abundance of helium given by Morrison & McCammon (1983). Assuming that the accretion disc extends all the way down to the last stable orbit at radius $R = 3R_S = 6GM/c^2$ without striking the neutron star, the efficiency is $\eta = 0.083$, and the Eddington limit corresponds to an accretion rate

$$\dot{M}_{\rm Edd} = 3.1 \times 10^{-8} \left(\frac{M}{M_\odot}\right) \, M_\odot \, {\rm yr}^{-1}. \quad (1)$$

These values are consistent with the Newtonian dynamics assumed by Shakura & Sunyaev (1973).

At a given radius $R$, the density in the accretion disc is taken to be uniform in the vertical direction (see Laor & Netzer 1989). Choosing dimensionless parameters

$$m = \frac{M}{M_\odot}, \quad f = \frac{\dot{M}}{\dot{M}_{\rm Edd}}, \quad {\rm and} \quad r = \frac{R}{R_S}, \quad (2)$$

the flux emerging from the surface of the disc is given by

$$F_0 = \frac{1.2 \times 10^{27} f \left(1 - \sqrt{3/r}\right)}{mr^3} \, {\rm erg \, cm}^{-2} \, {\rm s}^{-1}. \quad (3)$$

The density of the gas is given by

$$\rho_0 = \frac{3.4 \times 10^{-7} r^{3/2}}{\alpha m f^2 \left(1 - \sqrt{3/r}\right)^2} \, {\rm g \, cm}^{-3}, \quad (4)$$

and the half-thickness of the disc (measured in units of the Schwarzschild radius) is given by

$$h = \frac{H}{R_S} = 9.0 f \left(1 - \sqrt{3/r}\right). \quad (5)$$

A value $\alpha = 0.1$ is assumed for the viscosity parameter.

At each radius in the disc, the radiative transfer is treated using the Fokker-Planck/diffusion equation of Ross, Weaver & McCray (1978) in plane-parallel geometry,

$$\frac{\partial n}{\partial t} = 0 = \frac{N_e \sigma_T}{m_e c} \frac{1}{\varepsilon^2} \frac{\partial}{\partial \varepsilon} \left\{ \varepsilon^4 \left[ n + \left(kT + \frac{7\varepsilon^2}{10 m_e c^2}\right) \frac{\partial n}{\partial \varepsilon} \right] \right\}$$
$$+ \frac{\partial}{\partial z}\left(\frac{c}{3\kappa} \frac{\partial n}{\partial z}\right) + \frac{j_{\rm ff} h^3 c^3}{8\pi \varepsilon^3} - c\kappa_{\rm ff} n. \quad (6)$$

Here $n$ is the photon occupation number, $\varepsilon$ is the photon energy, $z$ is the vertical height within the disc, $N_e$ is the free-electron number density, $T$ is the gas temperature, $j_{\rm ff}$ is the free-free spectral emissivity, $\kappa$ is the total opacity (per volume) and $\kappa_{\rm ff}$ is the free-free absorption opacity. The first term on the right-hand side of equation (6) treats the full effects of incoherent Compton scattering, including the lowest order Klein-Nishina correction to the scattering rate. The remaining terms account for vertical diffusion of radiation as well as emission and absorption by free-free processes. The equation is solved in finite-difference form over a rectangular grid of $\varepsilon$ and $z$ values as described by Ross, Weaver & McCray (1978). The outer boundary condition for the radiation emerging from the surface of the disc has been discussed by Foster, Ross & Fabian (1986).

The inner boundary condition employed in solving equation (6) deserves consideration. At a given radius, the radiative transfer could be treated for the entire half-thickness of the disc; the inner boundary condition would then consist of setting the spectral flux equal to zero ($F_\nu = 0$). In order to reduce the computing time, however, we only treat a portion of the disc thickness as described by Ross, Fabian & Mineshige (1992). Care is taken that the "surface layer" treated is *effectively thick*, so that

$$\tau_* = \sqrt{3\tau_{\rm ff}(\tau_T + \tau_{\rm ff})} > 1, \quad (7)$$

even for high-energy photons. The spectral flux entering the surface layer from below is then assumed to satisfy the LTE radiative diffusion equation (see Rybicki & Lightman 1979),

$$F_\nu = -\frac{4\pi}{3\kappa} \frac{\partial B_\nu(T)}{\partial T} \frac{dT}{dz}, \quad (8)$$

where $B_\nu(T)$ is the Planck function. The LTE temperature at the base of the surface layer is estimated as described by Ross, Fabian & Mineshige (1992), while the local temperature gradient is chosen so that the total flux entering the surface layer is consistent with the total rate of dynamic heating in the gas beneath the surface layer. In this way, the *a priori* uncertainty in the temperature at the base of the surface layer has very little effect on the calculated emergent spectrum.

The occupation numbers, $n(\varepsilon, z)$, are relaxed until equation (6) is satisfied throughout the grid of $\varepsilon$ and $z$ values. During the relaxation, the local temperature, $T(z)$, is found by balancing the total heating rate due to dynamic heating, Compton scattering, and free-free absorption with the total cooling rate due to inverse Compton scattering and free-free emission. Thus completely self-consistent, non-LTE solutions are found for the radiation field and the temperature structure of the gas.

## 3  RESULTS

It should be noted that the parameter $f$ gives the fraction of the Eddington luminosity produced *solely* by the accretion disc. At least an equal luminosity must be produced in a thin boundary layer near the surface of the neutron star as the infalling gas loses its orbital kinetic energy (assuming that the neutron star is rotating slowly). Therefore, we limit our consideration to values $f < 1/2$. Sunyaev & Shakura (1986) have shown that the luminosity of the boundary layer can be as high as several times the luminosity of the accretion disc, depending on the actual radius of the neutron star. For $f < 1/2$, the inner accretion disc also remains geometrically thin ($h \ll r$) as assumed by Shakura & Sunyaev (1973).

### 3.1  Local spectrum at $r = 7$

First, we set $f = 0.3$ and consider the local spectrum emerging from the accretion disc at $r = 7$. This is chosen as a characteristic spectrum since the quantity $F_0 R^2$, which measures the contribution to the total disc luminosity, peaks at $r \approx 7$. For this model, the midplane of the disc is at a Thomson depth $\tau_T \approx 550$. To assure that high-energy photons are thermalized at the base of the surface layer, we treat the radiative transfer down to a Thomson depth $\tau_T \approx 275$. The calculated photon flux emerging from the surface of the disc is shown in Fig. 1. Also shown is a blackbody spectrum at the effective temperature $T_{\rm eff} = (F_0/\sigma)^{1/4} = 8.2 \times 10^6$ K. At low photon energies ($\varepsilon \lesssim 0.2$ keV), the calculated flux mimics a blackbody at the surface temperature, which is somewhat lower than $T_{\rm eff}$. At higher energies, however, the effects of Compton scattering become important. In the soft X-ray regime, the photon spectrum is nearly flat, while most of the flux emerges as a "Wien tail" in hard X-rays ($\varepsilon \gtrsim 2$ keV).

The hard X-ray tail results from photons escaping from deeper, hotter layers of the accretion disc. The competition between Compton scattering of higher-energy photons and inverse Compton scattering of lower-energy photons results in a Wien-law spectrum,

$$F_\nu \propto \nu^3 \exp(-h\nu/kT_{\text{Wien}}). \tag{9}$$

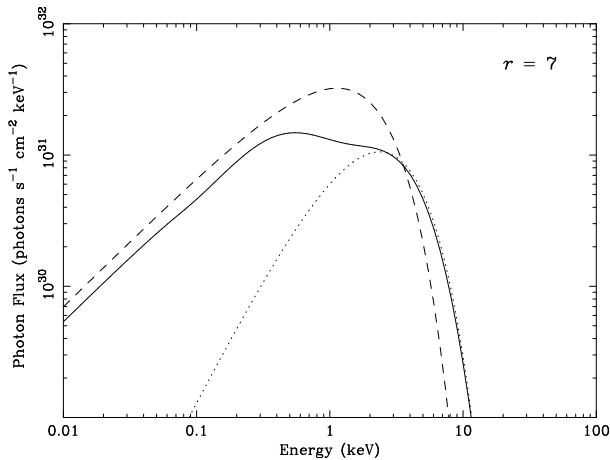

**Figure 1.** Calculated emergent spectrum (solid curve) at $r = 7$ for $f = 0.3$. Shown for comparison are the effective blackbody spectrum at $T_{\text{eff}} = 8.2 \times 10^6$ K (dashed curve) and a Wien-law spectrum at $T_{\text{Wien}} = 1.37 \times 10^7$ K with the same total energy flux (dotted curve).

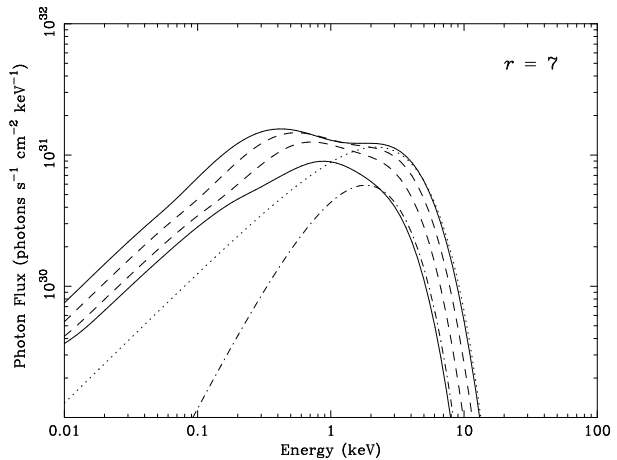

**Figure 2.** Calculated emergent spectra at $r = 7$ for $f = 0.1$ (lower solid curve), 0.2 (lower dashed curve), 0.3 (upper dashed curve), and 0.4 (upper solid curve). Also shown are a Wien-law spectrum (dash-dot curve) at $T_{\text{Wien}} = 1.06 \times 10^7$ K with the same total energy flux as the $f = 0.1$ model and the diluted blackbody spectrum (dotted curve) with $q = 1.8$ corresponding to the $f = 0.4$ model.

To estimate the Wien temperature, $T_{\text{Wien}}$, we examine the calculated temperature structure of the disc, seeking the Thomson depth for which the Compton $y$ parameter equals unity (see Rybicki & Lightman 1979):

$$y = \frac{4kT(\tau_{\text{T}})}{m_e c^2}\tau_{\text{T}}^2 = 1. \tag{10}$$

The resulting temperature should represent the lowest for which Comptonization is complete and, therefore, is appropriate for the Wien law. For the model with $f = 0.3$ and $r = 7$, equation (10) is satisfied at $\tau_{\text{T}} \approx 11$ where $T = 1.37 \times 10^7$ K. Assuming that the entire energy flux emerges as a Wien spectrum with this temperature, we get the result shown in Fig. 1. It is seen to be in excellent agreement with the high-energy tail of the actual emergent spectrum. Note that although high-energy photons are emitted at great depths where the temperature is higher, the Wien tail has a lower temperature corresponding to the smaller depth where $y = 1$. Without the effects of incoherent Compton scattering, the emergent spectrum would be even harder. As it is, the Wien tail greatly exceeds the effective blackbody spectrum for $\varepsilon \gtrsim 5$ keV.

The emergence of a Wien-law spectrum at high accretion rates was predicted by Shakura & Sunyaev (1973). Comptonization should be complete or "saturated" whenever the thermalization depth (where $\tau_* = 1$) for high-energy photons is greater than the Comptonization depth (where $y = 1$). In terms of our parameters $r$ and $f$, Shakura & Sunyaev predicted a Wien spectrum for

$$r \lesssim 93\alpha^{2/9}f^{2/3}. \tag{11}$$

For $\alpha = 0.1$ and $f = 0.3$, this gives $r \lesssim 25$, which accounts for nearly 3/4 of the total luminosity of the disc. Only farther out in the accretion disc did Shakura & Sunyaev predict that Comptonization would be incomplete, resulting in a "modified blackbody" spectrum.

Calculated emergent spectra at $r = 7$ for models with accretion rates corresponding to $f = 0.1$, 0.2, 0.3, and 0.4 are shown in Fig. 2. Even for $f = 0.1$, equation (11) predicts that Comptonization is complete for $r \lesssim 12$, accounting for half the total luminosity of the disc. For $f = 0.1$ and $r = 7$, our calculations show that equation (10) is satisfied at $\tau_{\text{T}} \approx 12$ where $T = 1.06 \times 10^7$ K. A Wien spectrum at this temperature containing the entire flux is shown in Fig. 2, and it is seen to be in reasonable agreement with the high-energy tail of the emergent spectrum.

A prominent Wien shoulder can be seen at $\varepsilon \approx 3$ keV for the highest accretion rate ($f = 0.4$) shown in Fig. 2. Shimura & Takahara (1995) used a "diluted blackbody" spectrum,

$$F_\nu = \frac{1}{q^4}\pi B_\nu(qT_{\text{eff}}), \tag{12}$$

to describe the Wien-law dominated portions of the local emergent spectra that they calculated. Here $q$ is an appropriate spectral hardening factor which, for high accretion rates, lies in the range $1.7 \lesssim q \lesssim 2.0$. In Fig. 2 we show the diluted blackbody spectrum for the $f = 0.4$ model assuming a value $q = 1.8$. It is seen to be in good agreement with the calculated spectrum for $\varepsilon \gtrsim 2$ keV.

Since we are interested in modelling high-luminosity LMXB accretion discs, we only consider high accretion rates ($f \geq 0.1$). Accurate treatment of the Wien-law tail is important in predicting the hard X-ray spectrum of such discs. Furthermore, the transition of the spectrum from a blackbody in the EUV to a Wien tail in hard X-rays can only be modelled accurately by a fully self-consistent treatment of the radiative transfer, as we do here.

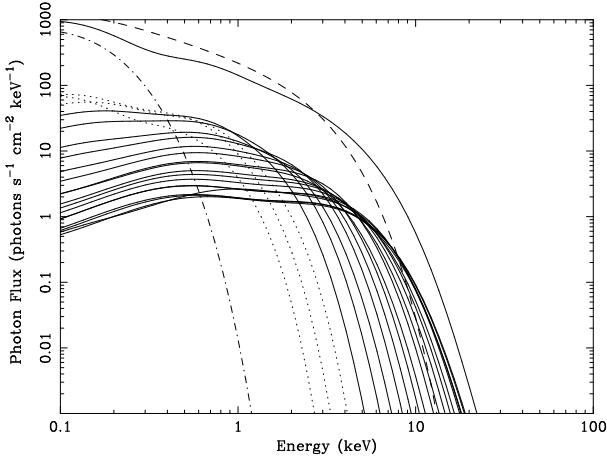

**Figure 3.** Summed spectra for an accretion disc with $f = 0.3$. Upper solid curve shows the total photon flux assuming a face-on disc at 1 kpc. Other curves show contributions to the total spectrum by different annuli assuming incoherent scattering for $3 < r < 70$ (solid curves), coherent scattering for $70 < r < 200$ (dotted curves), and blackbody spectra for $200 < r < 600$ (dash-dot curve). Also shown is the total spectrum if the entire disc is assumed to emit blackbody spectra (dashed curve).

### 3.2 Summed disc spectra

We divide the region $3 < r < 200$ of the accretion disc into 20 annuli that make comparable contributions to the total luminosity. The typical emergent spectrum for each annulus is calculated and multiplied by the area of the annulus to find the contribution to the spectral luminosity. The different contributions are added together, and the spectral photon flux is found assuming a face-on disc at an arbitrary distance of 1 kpc. The result for an accretion disc with $f = 0.3$ is shown in Fig. 3. For 17 annuli covering $3 < r < 70$, a complete treatment of incoherent Compton scattering is performed as described previously. For three annuli covering $70 < r < 200$, however, $y \ll 1$ at the thermalization depth. Therefore, we treat Compton scattering as being *coherent* for these calculations by dropping the Fokker-Planck term on the right-hand side of equation (6). Each coherent-scattering model is iterated until a self-consistent temperature is found for use in the inner boundary condition (equation 8). Finally, to give the spectrum a realistic behavior at very low photon energies, the spectrum for the region $200 < r < 600$ is found assuming that the disc emits an effective blackbody spectrum locally.

The resulting calculated disc spectrum should be very accurate for $\varepsilon \gtrsim 0.6$ keV. The dip at $\varepsilon \approx 0.4$ keV is probably somewhat exaggerated. Also shown in Fig. 3 is the total spectrum if the entire disc is assumed to emit an effective blackbody spectrum locally. The actual spectrum is seen to be much harder than that of a blackbody disc, with a greater spectral flux for $\varepsilon \gtrsim 3$ keV.

Fig. 4 shows the summed disc spectra in hard X-rays that we calculate for models with accretion rates corresponding to $f = 0.1, 0.2, 0.3,$ and $0.4$. We find that each total photon spectrum can be fit extremely well by a power law spectrum with an exponential cutoff,

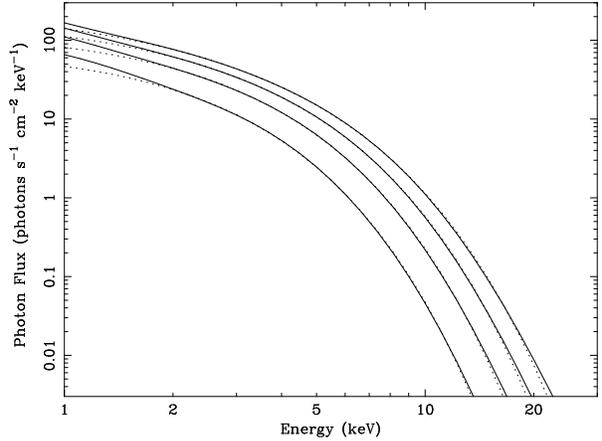

**Figure 4.** Summed spectra (solid curves) for accretion discs with $f = 0.1$ (lowest), 0.2, 0.3, and 0.4 (highest). Also shown (dotted curves) are fits made over the 2–20 keV band assuming cutoff power-law spectra.

**Table 1.** Cutoff power-law fits to summed disc spectra.

| $f$ | $\varepsilon_0$ (keV) | $\Gamma$ |
|---|---|---|
| 0.1 | 1.23 | $-0.19$ |
| 0.2 | 1.48 | $-0.09$ |
| 0.3 | 1.75 | 0.05 |
| 0.4 | 2.04 | 0.18 |

$$\frac{dN}{d\varepsilon} \propto (\varepsilon/\varepsilon_0)^{-\Gamma} \exp(-\varepsilon/\varepsilon_0). \tag{13}$$

The values of the parameters $\varepsilon_0$ and $\Gamma$ for the fits, made over the 2–20 keV band, are listed in Table 1. The values of $\Gamma$ are small, so the fits are close to simple exponential spectra. (Since we are fitting photon fluxes rather than energy fluxes, an exponential spectrum does *not* correspond to a single-temperature bremsstrahlung spectrum.) At low accretion rates, the $\Gamma$ values are negative, so the power-law factors *rise* with photon energy. As shown in Fig. 4, the fits are accurate descriptions of the calculated disc spectra over the decade $1.5 \lesssim \varepsilon \lesssim 15$ keV in photon energy.

### 3.3 Blurred disc spectra

The simple summed spectra shown in Figs. 3 and 4 are not completely accurate descriptions of the spectra observed by a distant observer. Even for a face-on disc, gravitational redshift and the transverse Doppler effect, which depend on the radius $r$, will modify the observed spectrum, making it softer. For an inclined disc, the flux emitted at a given local photon energy at radius $r$ will be spread over a range of observed energies by the Doppler effect. We take these effects into account to lowest order by using the method described by Chen, Halpern & Filippenko (1989) without the correction for light bending.

The resulting "blurred" disc spectrum for an accretion rate corresponding to $f = 0.3$ and an angle of inclination $i = 30°$ is shown in Fig. 5. Small wiggles in the spectrum are merely artifacts of interpolation between computed models. We find again that the observed hard X-ray spectrum is all of these spectra, cutoff power laws provide excellent fits over approximately a decade of hard X-ray energies. The parameters for the 2–20 keV fits are listed in Table 2. Since $-1/2 \lesssim \Gamma \lesssim 1/2$, all fitting functions are dominated by the exponential term. For a given accretion rate, both $\varepsilon_0$ and $\Gamma$ rise monotonically with increasing inclination. Similarly, both parameters rise monotonically with increasing accretion rate for a fixed angle of inclination.

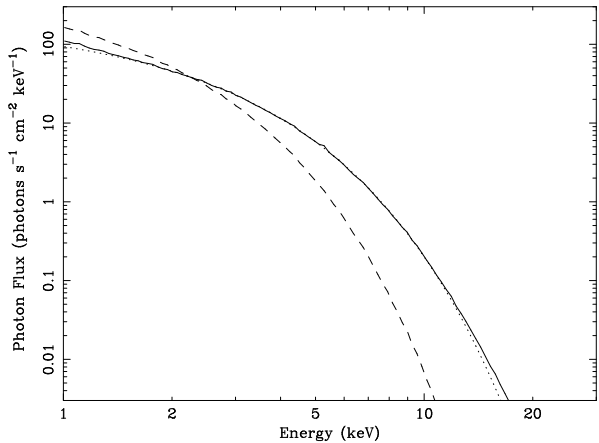

**Figure 5.** Observed hard X-ray spectrum (solid curve) for an accretion disc with $f = 0.3$ inclined at $i = 30°$, taking into account Doppler blurring and gravitational redshift. Dotted curve is a 2–20 keV fit assuming a cutoff power-law spectrum. Dashed curve shows the spectrum that would be observed for a blackbody disc.

**Table 2.** Cutoff power-law fits to blurred disc spectra.

| $f$ | $i$ | $\varepsilon_0$ (keV) | $\Gamma$ |
|---|---|---|---|
| 0.1 | 10° | 0.95 | −0.45 |
| " | 30° | 1.07 | −0.17 |
| " | 50° | 1.22 | 0.02 |
| 0.2 | 10° | 1.15 | −0.25 |
| " | 30° | 1.29 | −0.05 |
| " | 50° | 1.48 | 0.15 |
| 0.3 | 10° | 1.37 | −0.06 |
| " | 30° | 1.51 | 0.09 |
| " | 50° | 1.74 | 0.26 |
| 0.4 | 10° | 1.61 | 0.12 |
| " | 30° | 1.76 | 0.23 |
| " | 50° | 2.03 | 0.37 |

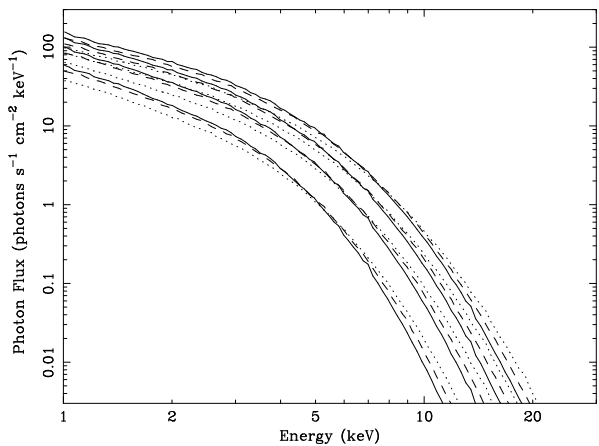

**Figure 6.** Observed hard X-ray spectra for accretion discs with $f = 0.1$ (lowest), 0.2, 0.3, and 0.4 (highest) inclined at $i = 10°$ (solid curves), 30° (dashed curves), and 50° (dotted curves).

well fit by a cutoff power-law spectrum. While $\Gamma$ is nearly unchanged, the cutoff energy is reduced to $\varepsilon_0 = 1.51$ keV since the spectrum is softer than the simple face-on summed spectrum. As shown in Fig. 5, the cutoff power-law fit is accurate over the decade of hard X-ray energies responsible for most of the luminosity of the accretion disc. For comparison, Fig. 5 also shows the blurred spectrum that would result if each point on the surface of the accretion disc emitted its effective blackbody spectrum. The observed spectrum is considerably harder than such a blackbody disc spectrum.

Blurred hard X-ray spectra for accretion rates corresponding to $f = 0.1$, 0.2, 0.3, and 0.4 are shown in Fig. 6 for inclinations of 10°, 30°, and 50°. For a given accretion rate, the lowest inclination gives the strongest flux at low photon energies, but at high energies Doppler blurring causes the strongest flux to occur at the highest inclination. For

## 4 DISCUSSION

We see that the observed hard X-ray photon spectrum for a luminous LMXB accretion disc should resemble a soft ($1 \lesssim \varepsilon_0 \lesssim 2$ keV) exponential decay multiplied by a shallow ($-1/2 \lesssim \Gamma \lesssim 1/2$) power law. If the magnetosphere disrupts the disc before it reaches $3R_S$, then the spectrum of the disc will be even softer. Of course, the accretion disc spectrum only represents one contribution to the total observed spectrum of a LMXB. Other contributions may come from a boundary layer at the neutron star's surface or magnetosphere and/or from a hot corona above the accretion disc (e.g., see Lamb 1991).

Fitting *EXOSAT* spectra of five bright low-mass X-ray binaries with a cutoff power law plus a blackbody spectrum, White et al. (1986) found higher values ($0.6 \lesssim \Gamma \lesssim 1.6$ and $3 \lesssim \varepsilon_0 \lesssim 6$ keV) for the power-law parameters. They assumed that the cutoff power law and blackbody represented the accretion disc and boundary layer, respectively. The discrepancy between their results and our models may in part be due to a single-temperature blackbody spectrum being a poor representation of the boundary layer and other emission regions of LMXB. The difference between observed total spectra and the accretion-disc spectra that we have calculated should give important spectral information about these poorly understood emitting regions. Further information may come from the time variability of the different components.

## ACKNOWLEDGEMENTS

RRR and ACF thank the College of the Holy Cross and the Royal Society, respectively, for support. RRR thanks the Institute of Astronomy for its hospitality.

**REFERENCES**

Chen K., Halpern J. P., Filippenko A. V., 1989, ApJ, 339, 742
Czerny B., Czerny M., Grindlay J. E., 1986, ApJ, 311, 241
Foster A. J., Ross R. R., Fabian A. C., 1986, MNRAS, 221, 409
Laor A., Netzer H., 1989, MNRAS, 238, 897
Lamb F. K., 1991, in Ventura J., Pines D., eds, Neutron Stars: Theory and Observation. Kluwer, Dordrecht, p. 445
Mitsuda K. et al., 1984, Pub. Astron. Soc. Japan, 36, 741
Morrison R., McCammon D., 1983, ApJ, 270, 119
Nannurelli M., Stella L., 1989, A&A, 226, 343
Ross R. R., Fabian A. C., Mineshige S., 1992, MNRAS, 258, 189
Ross R. R., Weaver R., McCray R., 1978, ApJ, 219, 292
Rybicki G. B., Lightman, A. P., 1979, Radiative Processes in Astrophysics. Wiley, New York
Shakura N. I., Sunyaev R. A., 1973, A&A, 24, 337
Shimura T., Takahara F., 1993, ApJ, 419, 78
Shimura T., Takahara F., 1995, ApJ, 445, 780
Sunyaev R. A., Shakura N. I., 1986, Sov. Astron. Lett., 12, 117
Taam R. E., Meszaros P., 1987, ApJ, 322, 329
White N. E., Peacock A., Hasinger G., Mason K. O., Manzo G., Taylor B. G., Branduardi-Raymont G., 1986, MNRAS, 218, 129
White N. E., Stella L., Parmar A. N., 1988, ApJ, 324, 363